\documentclass{article}

\usepackage[english]{babel}
\usepackage[utf8]{inputenc}
\usepackage{latexsym,amssymb,amstext,amsfonts,amsmath,graphicx,setspace,mathtools,bm}
\usepackage[colorinlistoftodos]{todonotes}
\usepackage{enumerate} 
\usepackage{paralist}
\usepackage{float}
\usepackage{subcaption}
\usepackage{array}
\usepackage{comment}
\usepackage{fullpage}
\usepackage{multirow}

\usepackage{tikz}
\usetikzlibrary{fit,positioning}
\usetikzlibrary{calc}

\tikzset{>=latex}

\usepackage{lineno}

\newcommand{\NewForRevision}[1]{#1}

\newcommand{\CapOpt}[1]{#1}

\definecolor{nice_blue}{RGB}{65, 105, 225}
\definecolor{nice_red}{RGB}{168, 34, 34}
\definecolor{IGCBlue}{HTML}{16197A}
\definecolor{dark_green}{RGB}{20, 110, 10}

\definecolor{graph_blue}{RGB}{144, 195, 212}
\definecolor{graph_purple}{RGB}{195, 144, 212}
\definecolor{graph_green}{RGB}{161, 212, 144}
\definecolor{graph_orred}{RGB}{212, 161, 144}

\begin{document}

{
\large

\noindent\textbf{Title}: Scanner Invariant Representations for Diffusion MRI Harmonization\\[1em]
\textbf{Authors:} Daniel Moyer (1,2), Greg Ver Steeg (2), Chantal M. W. Tax (3), Paul M. Thompson* (1)\\[1em]
\textbf{Institutions: }\\
(1) Imaging Genetics Center, Mark and Mary Stevens Institute for Neuroimaging and Informatics, Keck School of Medicine, University of Southern California, Los Angeles, CA, USA\\
(2) Information Sciences Institute, Marina del Rey, CA, USA \\
(3) CUBRIC, School of Psychology, Cardiff University, Cardiff, United Kingdom\\[1em]
\noindent \textbf{Word count: }5966\\
\textbf{Figure and table count: }10

*Corresponding author email: \texttt{pthomp@usc.edu}
}

\pagebreak

\title{Scanner Invariant Representations\\ for Diffusion MRI Harmonization}
%
%
%
%
%


\section*{Abstract}

\noindent\textbf{Purpose:} In the present work we describe the correction of diffusion-weighted MRI for site and scanner biases using a novel method based on invariant representation.\\

\noindent\textbf{Theory and Methods:} Pooled imaging data from multiple sources are subject to variation between the sources. Correcting for these biases has become very important as imaging studies increase in size and multi-site cases become more common. We propose learning an intermediate representation invariant to site/protocol variables, a technique adapted from information theory-based algorithmic fairness; by leveraging the data processing inequality, such a representation can then be used to create an image reconstruction that is uninformative of its original source, yet still faithful to underlying structures. To implement this, we use a deep learning method based on variational auto-encoders (VAE) to construct scanner invariant encodings of the imaging data.\\

\noindent\textbf{Results: } To evaluate our method, we use training data from the 2018 MICCAI Computational Diffusion MRI (CDMRI) Challenge Harmonization dataset. Our proposed method shows improvements on independent test data relative to a recently published baseline method on each subtask, mapping data from three different scanning contexts to and from one separate target scanning context.\\

\noindent\textbf{Conclusion: }As imaging studies continue to grow, the use of pooled multi-site imaging will similarly increase. Invariant representation presents a strong candidate for the harmonization of these data.

%

\pagebreak

\section{Introduction}

Observational conditions may vary strongly within a medical imaging study. Researchers are often aware of these conditions (e.g., scanner, site, technician, facility, etc.) but are unable to modify the experimental design to compensate, due to cost or geographic necessity. In magnetic resonance imaging (MRI), variations in scanner characteristics such as the magnetic field strength, scanner vendor, receiver coil hardware, applied gradient fields, or primary image reconstruction methods may have strong effects on collected data \cite{chen2014exploration,fortin2017harmonization,jovicich2006reliability}; multi-site studies in particular are subject to these effects \cite{hawco2018longitudinal,kelly2017widespread,magnotta2012multicenter,zavaliangos2018diffusion}. Data harmonization is the process of removing or compensating for this unwanted variation through post-hoc corrections. In the present work we focus on harmonization for diffusion MRI (dMRI), a modality known to have scanner/site biases \cite{correia2009looking,giannelli2010dependence,pagani2010intercenter,papinutto2013reproducibility,vollmar2010identical,white2009global,zhan2010does,zhan2013magnetic,zhan2012spatial} as well as several extra possible degrees of freedom with respect to protocol (e.g., angular resolution, $b$-values, gradient waveform choice, etc.).

Several prior methods approach diffusion MRI harmonization as a regression problem. Supervised image-to-image transfer methods have been proposed \cite{blumberg2018deeper,tanno2017bayesian}, while for the unsupervised case site effects are often modeled as covariate effects, either at a summary statistic level \cite{fortin2017harmonization,zavaliangos2018diffusion} or on the image directly \cite{mirzaalian2018multi}. All of these methods directly transform scans from one site/scanner context to another. Further, while all methods require paired scans to correctly validate their results (subjects or phantoms scanned on both target and reference scanners), supervised methods also require paired training data.  The collection of such data is expensive and difficult to collect at a large scale.


In this paper we instead frame the harmonization problem as an unsupervised image-to-image transfer problem\NewForRevision{, where harmonizing transformations may be learned without explicitly paired scans}. We propose that a subset of harmonization solutions may be found by learning scanner-invariant representations, i.e., representations of the images that are uninformative of which scanner the images were collected on. These representations and the mappings between them may then be manipulated to provide image reconstructions that are minimally informative of their original collection site. We thus provide an encoder/decoder method for learning mappings to and from invariant representations computationally. This method has several advantages over regression-based methods, including a practical implementation that does not require paired data, i.e., a traveling phantom as training input, and an extension to a multi-site case.

We demonstrate our proposed method on the MICCAI Computational Diffusion MRI challenge dataset \cite{ning2018muti,taxcross, zhu2018challenges}, showing substantial improvement compared to a recently published baseline method. We also introduce technical improvements to the training of neural architectures on diffusion-weighted data, and discuss the limitations and error modes of our proposed method.

\subsection{Relevant Prior Work}

Harmonization has been an acknowledged problem in MR imaging and specifically diffusion imaging for some time \cite{zhu2018challenges}. Numerous studies have noted significant differences in diffusion summary measures (e.g., fractional anisotropy; FA) between scanners and sites \cite{pagani2010intercenter,vollmar2010identical,white2009global}. Further protocol differences arise between sites due to limitations of the available scanners, unavoidable changes or upgrades in scanners or protocols, or when combining data retrospectively from multiple studies; effects of variations in scanning protocols on derived measures include effects of voxel size \cite{papinutto2013reproducibility}, $b$-values (the diffusion weightings used) \cite{correia2009looking,papinutto2013reproducibility}, and angular resolution or $q$-space sampling \cite{giannelli2010dependence,zhan2010does,zhan2013magnetic,zhan2012spatial} among other parameters. These problems were also examined by the MICCAI Computational Diffusion MRI 2017 and 2018 challenges \cite{ning2018muti,taxcross}, which held an open comparison of methods for a supervised (paired) task.

Most previously proposed harmonization methods have relied on forms of regression. Harmonization of summary statistics (voxel-wise or region-wise) include random/mixed-effect models \cite{zavaliangos2018diffusion} as well as the scale-and-shift random effects regression of ComBat \cite{fortin2017harmonization,zavaliangos2018diffusion}. This latter method was adapted from the genomics literature \cite{johnson2007adjusting}, and employs a variational Bayes scheme to learn model coefficients.

A more nuanced family of regression methods for diffusion imaging was recently introduced in a series of papers by Mirzaalian et al. \cite{mirzaalian2016inter,mirzaalian2018multi,mirzaalian2015harmonizing}. This was later analyzed empirically in Karayumak et al. \cite{karayumak2019retrospective}, which compared it against ComBat \cite{johnson2007adjusting} for summary statistics. This family of methods computes a power spectrum from a spherical harmonic (SH) basis, then generates a template from these images using multi-channel diffeomorphic mappings. The resulting template is used to compute spatial maps of average SH power spectra by scanner/protocol, which are then used in a scale regression on individual subjects. While these papers take a very different approach from our own, the resulting method has a very similar usage pattern and output. We compare our approach directly to this method.

In a supervised (paired) task, direct image-to-image transfer has been explored both in the harmonization context \cite{karayumak2018harmonizing,koppers2018spherical,ning2018muti} as well as the similar super-resolution context \cite{blumberg2018deeper,tanno2017bayesian}. This family of methods generally relies on high expressive-capacity function fitting (e.g., neural networks) to map directly between patches of pairs of images. This requires explicitly paired data, in that the same brains must be scanned at all sites. These methods perform well empirically, as tested by the CDMRI challenge \cite{taxcross}, but require paired data in the training set. Our proposed method does not require paired data to train; however, in our opinion, best practice validation still requires paired data in the (holdout) test-set.





\section{Theory}

Our goal is to map diffusion MRI scans from one scanner/site context to another, so that given an image from one site we could predict accurately what it would have looked like were it collected at another site. In order to do this, we construct an encoding function $q$ that takes each image $x$ to a corresponding vector $z$, and a conditional decoding function $p$ that takes each $z$ and a specified site $s$ back to an image $\hat{x}$ (the ``reconstruction'' of the original image). 

We further wish to remove trends and biases in $x$ that are informative of $s$ from the reconstruction $\hat{x}$, so that all data remapped to a given site $s'$ have the same bias (this is the harmonization task). In order to do so, it would be sufficient to constrain $z$, the intermediate representation, to be independent of $s$, denoted $z \perp s$. This is a hard constraint, and direct optimization of $q$ and $p$ subject to that constraint would be non-trivially difficult.

Instead, we choose to relax the constraint $z \perp s$ to the mutual information $I(z,s)$. Mutual information, taken from information theory, quantifies the amount of information shared between two variables e.g. $z$ and $s$. $I(z,s) = 0$ if and only if $z \perp s$, and so its minimization is a relaxation of our desired constraint. For a comprehensive reference on information theory, we refer the reader to Ch. 2 and Ch. 8 of Cover and Thomas \cite{cover2012elements}.

After relaxing the independence constraint to mutual information, we would like to optimize $q$ and $p$ so that $q(z|x)$ has minimal scanner-specific information, and so that $p(x|z)$ has minimal differences from the original data. We demonstrate one solution for doing this using a variational bound on $I(z,s)$, parameterizing $p$ and $q$ using neural networks. The underlying theory is explored in Moyer et al. \cite{NIPS2018_8122}, where it is used in the context of algorithmic fairness. We reproduce it here for clarity, and further reinterpret their theoretical results in the imaging harmonization context, adding our own data processing inequality interpretation of test-time remapping.


Learning the mapping $q$ does \emph{not} require matching pairs of data $(x,x')$ from pairs of sites $(s,s')$. 
Best practices in validation and testing \emph{do} require such data, but during training we can minimize $I(z,s)$ without having examples of the same subject collected on different scanners. This is due to our bound of $I(z,s)$ described in Eq. \ref{eq:IZS}, which is not reliant on inter-site correspondence. 

At test time we can manipulate this mapping to reconstruct images at a different site than they were originally collected at; we use this mapping as our harmonization tool. Again, by the data processing inequality, the amount of information these (new) reconstructed images contain about their original collection site is bounded by $I(z,s)$, which we explicitly minimize.

\subsection{Scanner Invariant Variational Auto-Encoders}
\label{subsec:inv_vae}

We wish to learn a mapping $q$ from data $x$ (associated with scanner $s$) to some latent space $z$ such that $z \perp s$, yet also where $z$ is maximally relevant to $x$. We start by relaxing $z\perp s$ to $I(z,s)$, and then bounding $I(z,s)$ (detailed demonstration in Appendix \ref{sec:app:bound}):
\begin{align}
I(z,s) \leq
\underbrace{- \mathbb{E}_{x,s,z\sim q}[ \log p(x|z,s)]}_{\text{Conditional Reconstruction}}
+ \underbrace{\mathbb{E}_{x}[~KL[~q(z|x)~\|~q(z)~]~  ]}_{\text{Compression}}
- \underbrace{\vphantom{\mathbb{E}_{x,q}[]}H(x|s)}_{\text{Const}} \label{eq:IZS}
\end{align}
where $q(z)$ is the empirical marginal distribution of $z$ under $q(z|x)$, the specified encoding which we control, and $p(x|z,s)$ is a variational approximation to the conditional likelihood of $x$ given $z$ and $s$ again under $q(z|x)$. Here, $KL$ denotes the Kullback-Leibler divergence and $H$ denotes Shannon entropy.

The bound in Eq.\ref{eq:IZS} has three components: a conditional reconstruction, a compressive divergence term, and a constant term denoting the conditional entropy of the scan given the scanner. We can interpret Eq.\ref{eq:IZS} as stating that the information in $z$ about $s$ is bounded above by uncertainty of $x$ given $z$ and $s$, plus a penalty on the information in $z$ and a constant representing the information $s$ has about $x$ overall.

Intuitively, this breakdown makes sense: if we reconstruct given $s$, and are otherwise compressing $z$, the optimal compressive $z$ has no information about $s$ for reconstruction; $q(z|x)$ can always remove information about $s$ without penalty, because the reconstruction term is handed that information immediately. Further, if $x$ is highly correlated with $s$, i.e. $H(x|s)$ is very low, then our bound will be worse.

We can now construct a variational encoding/conditional-decoding pair $q$ and $p$ which fits our variational bound of $I(z,s)$ nicely, and which also fits our overall goal of re-mapping $x$ accurately through $p(x|z,s)$.
Following Kingma and Welling \cite{kingma2013auto}, we use a generative log-likelihood as an objective:

\begin{align}
\max \log \mathbb{E}_{(x,s)}[p(x | s)]
\end{align}

Here however, we inject the conditional likelihood to match our bound for $I(z,s)$. This also fits our test-time desired Markov chain (with condition $z\perp s$) \NewForRevision{where $\hat{x}$ is the harmonized reconstruction at new site $s'$}:
\begin{align*}
s \rightarrow x \rightarrow z \rightarrow \hat{x} \leftarrow s'
\end{align*}
Following the original VAE derivation (again in Kingma and Welling), we can derive a similar VAE with $s$-invariant encodings by introducing the encoder $q(z|x)$:
\begin{align}
\log p(x|s) & = \log \int \frac{p(x,z|s)}{q(z|x)} q(z|x) dz = \log \mathbb{E}_{z\sim q}\left[ \frac{p(x,z|s)}{q(z|x)} \right]\\
& \geq \mathbb{E}_{z\sim q}[ \log p(x,z|s) - \log q(z|x) ]\\
& = \mathbb{E}_{z\sim q}[\log p(z|s) - \log q(z|x) +  \log p(x|z,s) ] .
\end{align}
We assume that the prior $p(z|s) = p(z)$, i.e., that the conditional prior is equal to the marginal prior over $z$. In the generative context, this would be a strong model mis-specification: if we believe that there truly are generating latent factors, it is unlikely that those factors would be independent of $s$. However, we are not in such a generative frame, and instead would like to find a code $z$ that is invariant to $s$, so it is reasonable to use a prior that also has this property.
Taking this assumption, we have
\begin{align}
\log p(x|s) & \geq -KL[~q(z|x)~\|~p(z)] + \mathbb{E}_{z\sim q}[\log p(x|z,s)]. \label{eq:first_term}
\end{align}
This is a conditional extension of the VAE objective from Kingma and Welling \cite{kingma2013auto}. Putting this objective together with the penalty term in Eq.\ref{eq:IZS}, we have the following variational bound on the combined objective (up to a constant):
\begin{align}
\mathbb{E}_{(x,s)}[\log P(x|s)] - \lambda I(z,s) \geq ~~~~~~~~~~~~~~~~~~& \nonumber \\ 
 \mathbb{E}_{(x,s)}[ -\underbrace{KL[q(z|x)\|p(z)]}_{\text{Div. from Prior}}
 - \underbrace{\lambda KL[q(z|x)\|q(z)]}_{\text{Div. from Marg.}} + &(1+\lambda) \underbrace{\mathbb{E}_{z\sim q}[\log p(x|z,s)]]}_{\text{Cond. Reconstruction}}. \label{eq:unsup_objective}
\end{align}

We use the negation of Eq.\ref{eq:unsup_objective} as the main loss term for learning $q$ and $p$, where we want to minimize the negative of the bound. As described in Higgins et al. \cite{higgins2017beta}, an additional parameter $\alpha$ may be multiplied with the divergence from the prior (the first term of Eq. \ref{eq:unsup_objective}) for further control over the VAE prior.


As we have it written in Eq. \ref{eq:unsup_objective}, the site variable $s$ has ambiguous dimension. For applications with only two sites, $s$ might be binary, while in the multi-site case, $s$ might be a one-hot vector\footnote{For a categorical variable with value $k$ out of $K$ possible values, its corresponding one-hot vector is a $K$-dimensional vector with zeros in every entry except for the $k^{th}$ entry, which is one.}. We conduct experiments for both in Sections \ref{sec:methods} and \ref{sec:results}. More complex $s$ values are also possible, but we do not explore them in this paper.


\subsection{Diffusion-space Error Propagation from SH representations}

A convenient representation for diffusion-weighted MRI is the spherical harmonics (SH) basis \cite{mirzaalian2015harmonizing}. These provide a countable set of basis functions from the sphere to and from which projection is easy and often performed (e.g., in graphics). In this paper, our input data and the reconstruction error is computed with respect to the SH coefficients. However, for the eventual output, the data representation that we would like to use is \emph{not} in this basis, but in the original image representation which is conditional on a set of gradient vectors (b-vectors). These vectors are in general different for each subject due to spatial positioning and motion, and often change in number between sites/protocols. Rigid transformation and alignment of scan data, used in many pre-processing steps, also change vector orientation. While the $\ell_2$ function norm is preserved under projection to the SH basis (i.e., asymptotically SH projection is an isomorphism for $\ell_2$), this is not the case for a norm on general finite sets of vectors.

To correct for this, we construct a projection matrix from the shared continuous SH basis to\NewForRevision{ the diffusion gradient directions}. This projection can then be used in conjunction with decoder output $p(x|z,s)$ to map output SH coefficients to the original subject-specific $b$-vector representation. We allow each $b_0$ channel to ``pass through'' the projection (mapped as identity), as they are without orientation. While we use the SH representation for both input and reconstruction (to leverage our invariance results), we augment the loss function from Eq. \ref{eq:IZS} with a ``real-space'' loss function, the reconstruction loss in each subject's original domain. This encourages the overall loss function to be faithful to our use-case in the original image space.

\section{Methods}
\label{sec:methods}

\subsection{Computational Implementation}

\begin{figure}[h!]
\centering
\begin{subfigure}{1.0\textwidth}
\centering
\begin{tikzpicture}[x=0.6cm,y=0.6cm]

\filldraw[fill=gray!33!white, draw=black] (0,-2.25) rectangle (1,2.25);

\filldraw[fill=gray!50!white, draw=black,label=z] (5,-0.5) rectangle (6,1.5);
\filldraw[fill=red!33!white, draw=black,label=z] (5,-1.5) rectangle (6,-0.5);


\filldraw[fill=gray!33!white, draw=black] (10,-2.25) rectangle (11,2.25);

\filldraw[fill=gray!33!white, draw=black, dashed] (10,-0.5) rectangle (11,0.5);

\filldraw[fill=blue!33!white, draw=black] (13.5,0) rectangle (14.5,1.5);
\draw[thick,blue] (11,0.5) -- (13.5,1.5);
\draw[thick,blue] (11,-0.5) -- (13.5,0);


\filldraw[fill=dark_green!33!white, draw=black] (13.5,-2) rectangle (14.5,-0.5);
\draw[thick,dark_green] (11,2.25) -- (13.5,-0.5);
\draw[thick,dark_green] (11,-2.25) -- (13.5,-2);

\node[blue] (latent_code) at (12.2,0.25) { {\tiny  SH $\rightarrow$ Subj}};

\node[dark_green] (latent_code) at (12.25,-1.25) {\tiny Adversary};

\draw[thick] (1,-2.25) -- (5,-0.5);
\draw[thick] (1,2.25) -- (5,1.5);
\node (qzx) at (3,-0.15) {$q(z|x)$};
\node (qzx) at (3,0.55) {Encoder};

\node (z) at (5.5,0.5) {$z$};
\node (c) at (5.5,-1) {$s$};


\node (pxz) at (8.25,0.7) {Conditional};
\node (pxz) at (8.25,-0.0) {Decoder};

\node (pxz) at (8.25,-0.7) {$p(x|z,s)$};
\draw[thick] (6,1.5) -- (10,2.25);
\draw[thick] (6,-1.5) -- (10,-2.25);

\node (xrecon) at (10.5,0) {$\hat{x}$};
\node (x) at (0.5,0) {$x$};

\node [rotate=90] (xtext) at (-1,0) {{\footnotesize Input Patch (SH) }};

\node (xtext) at (-0.75,-3) {{\tiny Hidden Units}};
\node (xtext) at (-0.75,-3.4) {{\tiny Per Layer}};


\node (q-arch) at (2.7,-3) {$\underbrace{\phantom{\hspace{7em}}}_{256~\rightarrow~128~\rightarrow ~64}$};
\node (z-arch) at (5.5,-3) {$\underbrace{\phantom{\hspace{2em}}}_{\rightarrow ~\text{32}~ \rightarrow}$};
\node (q-arch) at (8.3,-3) {$\underbrace{\phantom{\hspace{7em}}}_{64~\rightarrow~128~\rightarrow ~256}$};

\node (q-arch) at (12.5,-3) {\color{dark_green} $\underbrace{\phantom{\hspace{4em}}}_{32~\rightarrow~32}$};


\draw[->,thick] (5.5,1.5) |- node[] {} ++(6,1.5) |- (15.5,3) {};

\node (Lcomp) at (18,3) {$\mathcal{L}_{marg}$ and $\mathcal{L}_{prior}$};

\draw[->,thick] (11,1.85) -- (15.5,1.85) {};
\node (Lcomp) at (16.5,1.85) {$\mathcal{L}_{recon}$};

\draw[->,thick] (14.5,0.75) -- (15.5,0.75) {};
\node (Lcomp) at (16.35,0.65) {$\mathcal{L}_{proj}$};

\draw[->,thick] (14.5,-1.25) -- (15.5,-1.25) {};
\node (Lcomp) at (16.25,-1.25) {$\mathcal{L}_{adv}$};

\end{tikzpicture}
    \caption{Diagram of network configuration and losses.}
    \label{fig:net}
\end{subfigure}\\[1em]
\begin{subfigure}{1.0\textwidth}
\centering
\begin{minipage}{0.49\textwidth}
\begin{tikzpicture}[x=0.6cm,y=0.6cm]

\def\cdelta{4.25};

\def\clshim{0.25};
\def\crshim{0};
\def\bboxshim{0.25};

\filldraw[fill=black!05!white, draw=black] (-\bboxshim,\bboxshim) rectangle (2*\cdelta + 3.5 + \bboxshim,-8.5 - \bboxshim - 1.0);
\node (train_text) at (\cdelta + 1.75,\bboxshim + 0.5) {\textbf{Training Configuration}};

\node (left_train_grad) at (0,-8.5 - 0.75) {};
\node (right_train_grad) at (2*\cdelta + 1.5,-8.5 - 0.75) {$\mathcal{L}=\|x - \hat{x} \| +...$};

\path [->] (right_train_grad)
  edge[thick,out=180,in=0]
  node[midway, fill=black!05!white] {Backpropagation}  (left_train_grad);
  
\node[rotate=90] (subj_a_text) at (-0.75,-1.25) {Subj $1$};
\node[rotate=90] (subj_b_text) at (-0.75,-4.25) {Subj $2$};
\node[rotate=90] (subj_c_text) at (-0.75,-7.25) {Subj $3$};

\filldraw[fill=graph_blue!40!white, draw=black] (0,0) rectangle (3.5,-2.5);
\node (scan_a) at (1.45,-1.25) {\includegraphics[width=1.5cm]{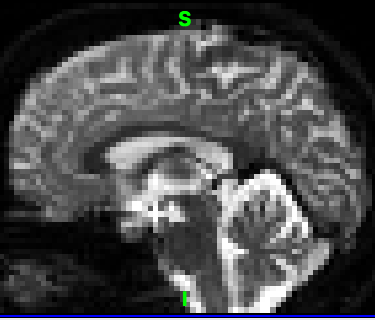}};
\node[rotate=90] (scan_a_text) at (3.125,-1.25) {Input $x$};
\node (scan_a_pt) at (3.5,-1.25) {};

\filldraw[fill=graph_orred!40!white, draw=black] (0,-3) rectangle (3.5,-5.5);
\node (scan_b) at (1.45,-4.25) {\includegraphics[width=1.5cm]{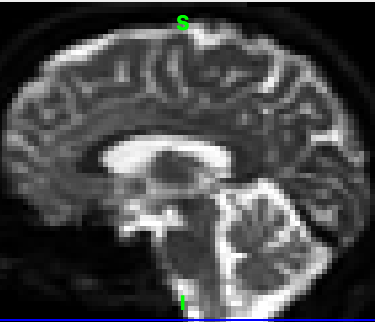}};
\node[rotate=90] (scan_b_text) at (3.125,-4.25) {Input $x$};
\node (scan_b_pt) at (3.5,-4.25) {};

\filldraw[fill=graph_green!40!white, draw=black] (0,-6) rectangle (3.5,-8.5);
\node (scan_c) at (1.45,-7.25) {\includegraphics[width=1.5cm]{intro/scan_b.png}};
\node[rotate=90] (scan_b_text) at (3.125,-7.25) {Input $x$};
\node (scan_c_pt) at (3.5,-7.25) {};


\filldraw[fill=black!20!white, draw=black] (\cdelta + \clshim,0) rectangle (\cdelta + 3.25 - \crshim,-1.2);
\node (inv_rep_text) at (\cdelta + 1.75, -0.6) {Siteless $z$};

\filldraw[fill=black!20!white, draw=black] (\cdelta + \clshim,-3) rectangle (\cdelta + 3.25 - \crshim,-4.2);
\node (inv_rep_text) at (\cdelta  + 1.75, -3.6) {Siteless $z$};

\filldraw[fill=black!20!white, draw=black] (\cdelta + \clshim,-6) rectangle (\cdelta + 3.25 - \crshim,-7.2);
\node (inv_rep_text) at (\cdelta + 1.75, -6.6) {Siteless $z$};

\node (inv_rep_pt_1) at (\cdelta + \clshim, -0.6) {};
\node (inv_rep_pt_2) at (\cdelta + \clshim, -3.6) {};
\node (inv_rep_pt_3) at (\cdelta + \clshim, -6.6) {};

\node (inv_rep_pt_1_out) at (\cdelta + 3.25 - \crshim, -0.6) {};
\node (inv_rep_pt_2_out) at (\cdelta + 3.25 - \crshim, -3.6) {};
\node (inv_rep_pt_3_out) at (\cdelta + 3.25 - \crshim, -6.6) {};

\filldraw[fill=black!20!white, draw=black] (\cdelta + \clshim,-1.5) rectangle (\cdelta + 3.25 - \crshim,-2.5);
\node (inv_rep_text_2) at (\cdelta + 1.375, -2.0) {Site $s$};
\filldraw[fill=graph_blue!80!white, draw=black] (\cdelta + 2.625, -2.0) circle (0.3);

\filldraw[fill=black!20!white, draw=black] (\cdelta + \clshim,-4.5) rectangle (\cdelta + 3.25 - \crshim,-5.5);
\node (inv_rep_text_2) at (\cdelta + 1.375, -5.0) {Site $s$};
\filldraw[fill=graph_orred!80!white, draw=black] (\cdelta + 2.625, -5.0) circle (0.3);

\filldraw[fill=black!20!white, draw=black] (\cdelta + \clshim,-7.5) rectangle (\cdelta + 3.25 - \crshim,-8.5);
\node (inv_rep_text_3) at (\cdelta + 1.375, -8.0) {Site $s$};
\filldraw[fill=graph_green!80!white, draw=black] (\cdelta + 2.625, -8.0) circle (0.3);

\node (osite_pt_1) at (\cdelta + 3.25 - \crshim, -2.0) {};
\node (osite_pt_2) at (\cdelta + 3.25 - \crshim, -5.0) {};
\node (osite_pt_3) at (\cdelta + 3.25 - \crshim, -8.0) {};

\path [->] (scan_a_pt) edge[thick] (inv_rep_pt_1);
\path [->] (scan_b_pt) edge[thick] (inv_rep_pt_2);
\path [->] (scan_c_pt) edge[thick] (inv_rep_pt_3);

\filldraw[fill=graph_blue!40!white, draw=black] (2*\cdelta,0) rectangle (2*\cdelta + 3.5,-2.5);
\node (scan_a) at (2*\cdelta + 1.45,-1.25) {\includegraphics[width=1.5cm]{intro/scan_a.png}};
\node[rotate=90] (scan_a_text) at (2*\cdelta + 3,-1.25) {Recon $\hat{x}$};
\node (scan_a_pt_out) at (2*\cdelta,-1.25) {};

\filldraw[fill=graph_orred!40!white, draw=black] (2*\cdelta,-3) rectangle (2*\cdelta + 3.5,-5.5);
\node (scan_a) at (2*\cdelta + 1.45,-4.25) {\includegraphics[width=1.5cm]{intro/scan_b.png}};
\node[rotate=90] (scan_a_text) at (2*\cdelta + 3,-4.25) {Recon $\hat{x}$};
\node (scan_b_pt_out) at (2*\cdelta,-4.25) {};

\filldraw[fill=graph_green!40!white, draw=black] (2*\cdelta,-6) rectangle (2*\cdelta + 3.5,-8.5);
\node (scan_a) at (2*\cdelta + 1.45,-7.25) {\includegraphics[width=1.5cm]{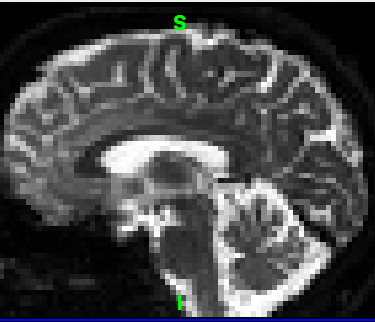}};
\node[rotate=90] (scan_a_text) at (2*\cdelta + 3,-7.25) {Recon $\hat{x}$};
\node (scan_c_pt_out) at (2*\cdelta,-7.25) {};

\path [->] (inv_rep_pt_1_out) edge[thick] (scan_a_pt_out);
\path [->] (inv_rep_pt_2_out) edge[thick] (scan_b_pt_out);
\path [->] (inv_rep_pt_3_out) edge[thick] (scan_c_pt_out);

\path [->] (osite_pt_1) edge[thick] (scan_a_pt_out);
\path [->] (osite_pt_2) edge[thick] (scan_b_pt_out);
\path [->] (osite_pt_3) edge[thick] (scan_c_pt_out);

\end{tikzpicture}

\end{minipage}
\begin{minipage}{0.49\textwidth}
\begin{tikzpicture}[x=0.6cm,y=0.6cm]

\def\cdelta{4.25};

\def\clshim{0.25};
\def\crshim{0};
\def\bboxshim{0.25};

\filldraw[fill=black!05!white, draw=black] (-\bboxshim,\bboxshim) rectangle (2*\cdelta + 3.5 + \bboxshim,-8.5 - \bboxshim - 1.0);
\node (train_text) at (\cdelta + 1.75,\bboxshim + 0.5) {\textbf{Testing Configuration}};

\node (left_train_grad) at (0,-8.5 - 0.75) {};
\filldraw[fill=black!20!white, draw=red] (0.5-0.25,-8.5 - 0.75 + 0.25 + 0.1) rectangle (0.5+0.25,-8.5 - 0.75 - 0.25 + 0.1);
\node (bot_text_1) at (3.5,-8.5 - 0.75) {Selected Target Site};

\filldraw[fill=black!20!white, draw=blue] (6.5-0.25,-8.5 - 0.75 + 0.25 + 0.1) rectangle (6.5+0.25,-8.5 - 0.75 - 0.25 + 0.1);
\node (bot_text_2) at (9.5,-8.5 - 0.75) {Harmonized Output};

\node[rotate=90] (subj_a_text) at (-0.75,-1.25) { };
\node[rotate=90] (subj_b_text) at (-0.75,-4.25) { };
\node[rotate=90] (subj_c_text) at (-0.75,-7.25) { };

\filldraw[fill=graph_blue!40!white, draw=black] (0,0) rectangle (3.5,-2.5);
\node (scan_a) at (1.45,-1.25) {\includegraphics[width=1.5cm]{intro/scan_a.png}};
\node[rotate=90] (scan_a_text) at (3.125,-1.25) {Input $x$};
\node (scan_a_pt) at (3.5,-1.25) {};

\filldraw[fill=graph_orred!40!white, draw=black] (0,-3) rectangle (3.5,-5.5);
\node (scan_b) at (1.45,-4.25) {\includegraphics[width=1.5cm]{intro/scan_b.png}};
\node[rotate=90] (scan_b_text) at (3.125,-4.25) {Input $x$};
\node (scan_b_pt) at (3.5,-4.25) {};

\filldraw[fill=graph_green!40!white, draw=black] (0,-6) rectangle (3.5,-8.5);
\node (scan_c) at (1.45,-7.25) {\includegraphics[width=1.5cm]{intro/scan_b.png}};
\node[rotate=90] (scan_b_text) at (3.125,-7.25) {Input $x$};
\node (scan_c_pt) at (3.5,-7.25) {};


\filldraw[fill=black!20!white, draw=black] (\cdelta + \clshim,0) rectangle (\cdelta + 3.25 - \crshim,-1.2);
\node (inv_rep_text) at (\cdelta + 1.75, -0.6) {Siteless $z$};

\filldraw[fill=black!20!white, draw=black] (\cdelta + \clshim,-3) rectangle (\cdelta + 3.25 - \crshim,-4.2);
\node (inv_rep_text) at (\cdelta  + 1.75, -3.6) {Siteless $z$};

\filldraw[fill=black!20!white, draw=black] (\cdelta + \clshim,-6) rectangle (\cdelta + 3.25 - \crshim,-7.2);
\node (inv_rep_text) at (\cdelta + 1.75, -6.6) {Siteless $z$};

\node (inv_rep_pt_1) at (\cdelta + \clshim, -0.6) {};
\node (inv_rep_pt_2) at (\cdelta + \clshim, -3.6) {};
\node (inv_rep_pt_3) at (\cdelta + \clshim, -6.6) {};

\node (inv_rep_pt_1_out) at (\cdelta + 3.25 - \crshim, -0.6) {};
\node (inv_rep_pt_2_out) at (\cdelta + 3.25 - \crshim, -3.6) {};
\node (inv_rep_pt_3_out) at (\cdelta + 3.25 - \crshim, -6.6) {};

\filldraw[fill=black!20!white, draw=red] (\cdelta + \clshim,-1.5) rectangle (\cdelta + 3.25 - \crshim,-2.5);
\node (inv_rep_text_2) at (\cdelta + 1.375, -2.0) {Site $s'$};
\filldraw[fill=graph_green!80!white, draw=black] (\cdelta + 2.625, -2.0) circle (0.3);

\filldraw[fill=black!20!white, draw=red] (\cdelta + \clshim,-4.5) rectangle (\cdelta + 3.25 - \crshim,-5.5);
\node (inv_rep_text_2) at (\cdelta + 1.375, -5.0) {Site $s'$};
\filldraw[fill=graph_green!80!white, draw=black] (\cdelta + 2.625, -5.0) circle (0.3);

\filldraw[fill=black!20!white, draw=red] (\cdelta + \clshim,-7.5) rectangle (\cdelta + 3.25 - \crshim,-8.5);
\node (inv_rep_text_3) at (\cdelta + 1.375, -8.0) {Site $s'$};
\filldraw[fill=graph_green!80!white, draw=black] (\cdelta + 2.625, -8.0) circle (0.3);

\node (osite_pt_1) at (\cdelta + 3.25 - \crshim, -2.0) {};
\node (osite_pt_2) at (\cdelta + 3.25 - \crshim, -5.0) {};
\node (osite_pt_3) at (\cdelta + 3.25 - \crshim, -8.0) {};

\path [->] (scan_a_pt) edge[thick] (inv_rep_pt_1);
\path [->] (scan_b_pt) edge[thick] (inv_rep_pt_2);
\path [->] (scan_c_pt) edge[thick] (inv_rep_pt_3);

\filldraw[fill=graph_green!40!white, draw=blue] (2*\cdelta,0) rectangle (2*\cdelta + 3.5,-2.5);
\node (scan_a) at (2*\cdelta + 1.45,-1.25) {\includegraphics[width=1.5cm]{intro/scan_c.png}};
\node[rotate=90] (scan_a_text) at (2*\cdelta + 3,-1.25) {Recon $\hat{x}$};
\node (scan_a_pt_out) at (2*\cdelta,-1.25) {};

\filldraw[fill=graph_green!40!white, draw=blue] (2*\cdelta,-3) rectangle (2*\cdelta + 3.5,-5.5);
\node (scan_a) at (2*\cdelta + 1.45,-4.25) {\includegraphics[width=1.5cm]{intro/scan_c.png}};
\node[rotate=90] (scan_a_text) at (2*\cdelta + 3,-4.25) {Recon $\hat{x}$};
\node (scan_b_pt_out) at (2*\cdelta,-4.25) {};

\filldraw[fill=graph_green!40!white, draw=blue] (2*\cdelta,-6) rectangle (2*\cdelta + 3.5,-8.5);
\node (scan_a) at (2*\cdelta + 1.45,-7.25) {\includegraphics[width=1.5cm]{intro/scan_c.png}};
\node[rotate=90] (scan_a_text) at (2*\cdelta + 3,-7.25) {Recon $\hat{x}$};
\node (scan_c_pt_out) at (2*\cdelta,-7.25) {};

\path [->] (inv_rep_pt_1_out) edge[thick] (scan_a_pt_out);
\path [->] (inv_rep_pt_2_out) edge[thick] (scan_b_pt_out);
\path [->] (inv_rep_pt_3_out) edge[thick] (scan_c_pt_out);

\path [->] (osite_pt_1) edge[thick] (scan_a_pt_out);
\path [->] (osite_pt_2) edge[thick] (scan_b_pt_out);
\path [->] (osite_pt_3) edge[thick] (scan_c_pt_out);

\end{tikzpicture}
\end{minipage}
    \caption{Diagram of training and testing schema for the proposed method.}
    \label{fig:train-test}
\end{subfigure}
\caption{
The \textbf{top} diagram describes the architecture of our method. The network is composed of an encoder branch $q(z|x)$ (at top left), a conditional decoder $p(x|z,s)$, as well as two augmenting losses. The DWI-space reconstruction loss in {\color{blue} blue} is computed using the injected subject-specific projection matrix (from SH to b-vector representation). The patch adversary in {\color{dark_green} green} attempts to predict whether a reconstructed patch is originally from a given site (``remapped'' vs. ``real'' patches). At test time the {\color{red} $s$ site id} is manipulated to map data onto one specific site.
The \textbf{bottom} diagram describes the training and testing schema for the proposed method, in the left and right boxes respectively. Site bias is represented by the differing colors. In both configurations, these are mapped to a site-invariant space $z$, the colorless center column of both boxes. The remaining (site-independent) information can then be reconstructed into an image, given a site. In the training configuration data are remapped to their original site, and the loss calculated from Eq. \ref{eq:loss} (secondary loss terms omitted from figure). Weights are then trained using the derivative of the loss (backpropagation \cite{rumelhart1985learning}). In the testing configuration, the data are mapped to the selected site $s'$. The outputs in the right-hand column of the right box have bounded mutual information about their original site, vanishing with respect to the loss function. 
}
\end{figure}

We parameterize $q$ and $p$ using neural networks, fitting their parameters by mini-batch gradient-based optimization. The loss in Eq. \ref{eq:unsup_objective} is defined generally, and invariant representations may be learned using many different function parameterizations. However, the flexibility of neural networks as function approximators make them ideal for this application. We apply these networks to small image patches, concatenating patch-wise outputs to create harmonized images. The overall architecture is shown in Figure \ref{fig:net}, and the training and testing configurations are diagrammed in Figure \ref{fig:train-test}, with exact parameters given in Section \ref{sec:methods}. We discuss the use of patches and its relative advantages and drawbacks in Section \ref{sec:disc}. \NewForRevision{Notably, in Figure \ref{fig:train-test} each sample consists of a single unpaired patch, and batches of data consist of patches and protocol identifiers (one-hot vectors). As diagrammed on the right-hand side of Figure \ref{fig:train-test}, protocol identifiers are manipulated at test time to produce harmonized reconstructions.}

Our primary reconstruction loss is computed in the SH domain with respect to the entire patch. We then add a secondary loss function for the center voxel based on the SH-to-DWI projection, and an adversarial loss which attempts to predict which scanner/protocol each reconstructed patch is from (seen at the right of Figure \ref{fig:net}). We added this branch in order to provide additional information toward keeping remapped patches ``reasonable'' when remapping to new sites; this prediction can be performed without explicit pairing of patches. Our loss function is then, in abstract,
\begin{align}
    \mathcal{L} = \mathcal{L}_{recon} + \alpha\mathcal{L}_{prior} + \beta\mathcal{L}_{proj} - \gamma\mathcal{L}_{adv} - \lambda I(z,s)\label{eq:loss}
\end{align}
where $\mathcal{L}_{recon}$ is SH reconstruction loss (using MSE), $\mathcal{L}_{proj}$ is the DWI space loss, and $\mathcal{L}_{adv}$ is the adversarial loss on the SH reconstruction, with three hyper parameters controlling trade-offs between objectives. This loss function trivially extends from the single-site case (one target site to/from one base site) to a multi-site case, where $s$ is categorical.

We use a standard adversarial training scheme for defining and minimizing $\mathcal{L}_{adv}$ (see e.g. Chapter 7.13 of Goodfellow, Bengio, and Courville \cite{goodfellow2016deep}). The adversarial loss $\mathcal{L}_{adv}$ is the softmax cross-entropy loss of a secondary ``adversary'' network, shown in green in Figure \ref{fig:net}. We alternate between optimizing the primary network (minimizing Eq. \ref{eq:loss}), and the adversary (minimizing $\mathcal{L}_{adv}$).

We optimize these networks by differentiating the loss functions (Eq. \ref{eq:loss} and $\mathcal{L}_{adv}$) with respect to the network weights (i.e. backpropagation \cite{rumelhart1985learning}) and then using the Adam optimizer \cite{kingma2014adam}, which is a first order optimization method. Optimization is undertaken using mini-batches. To compute gradients of the divergences in Eq. \ref{eq:unsup_objective} efficiently, we use the re-parameterization trick of Kingma and Welling \cite{kingma2013auto}, using both a diagonal Gaussian conditional $q(z|x)$ and a Gaussian prior $p(z)$. We also use the closed form bound for $KL[q(z|x)\|q(z)]$ from Moyer et al. \cite{NIPS2018_8122}.


\subsection{Data and Pre-processing}

To evaluate our method, we use the 15 subjects from the 2018 CDMRI Challenge Harmonization dataset \cite{tax2019cross,taxcross}. These subjects were imaged on two different scanners: a 3 T GE Excite-HD ``Connectom'' and a 3 T Siemens Prisma scanner. For each scanner, two separate protocols were collected, one of which matches between the scanners at a low resolution, and another which does not match at a high resolution. This results in four different ``site'' combinations, for which all subjects were scanned, resulting in forty different acquisitions (10 subjects, 2 scanners, 2 protocols each). We split this into 9 training subjects, 1 validation subject, and 5 held out-test subjects.

The low resolution matching protocol had an isotropic spatial resolution of 2.4 mm with 30 gradient directions ($\text{TE} = 89~ms$, $\text{TR}=7200~ms$)  at two shells $b=1200,3000~\frac{s}{mm^2}$, as well as a minimum of 4 $b_0$ acquisitions, at least one of these with reverse phase encoding. These volumes were then corrected for EPI distortions, subject motion, and eddy current distortions using FSL's TOPUP/eddy \cite{andersson2003correct,andersson2016integrated}. Subjects from the ``Connectom'' scanner were then registered to the ``Prisma'' scanner using a affine transformation, fit to a co-temporally acquired T1-weighted image volume (previously registered to each corresponding FA volume). The $b$-vectors were then appropriately rotated. In the case of the ``Connectom'' scanner, geometric distortions due to gradient non-linearities were corrected for using in-house software \cite{glasser2013minimal,rudrapatna2018can}. The high resolution protocols are identical in pre-processing to their low resolution counterparts, but have isotropic voxel sizes of $1.5$ mm ($\text{TE} = 80~ms$, $\text{TR}=4500~ms$) and 1.2 mm ($\text{TE} = 68~ms$, $\text{TR}=5400~ms$) for ``Prisma'' and ``Connectom'' scanners respectively, each with 60 gradient directions per shell, same b-shell configurations ($b=1200,3000\frac{s}{mm^2}$). We downsample the spatial resolution of the high resolution scans to $2.4$ mm isotropic to test the multi-task method, but keep the angular resolution differences.  To simplify notation, we refer to the four scanner/protocol combinations by their scanner make and number of gradient directions: Prisma 30, Prisma 60, Connectom 30, and Connectom 60.

All scans were masked for white matter tissue. This was done in order to focus our analysis on the tissue most commonly assessed using diffusion MRI (see e.g. for a review \cite{assaf2008diffusion}). We map each of these scans to an $8^{th}$-order SH representation for input into our method, but retain the original domain for training outputs. We use the minimal $\ell_2$ weighted solution in the case of under-determined projections, which corresponds with the SVD solution (using the pseudo-inverse). This is well-defined, unlike direct projection. 

\subsection{Experimental Protocol}
\label{subsec:protocol}

The original CDMRI 2018 challenge \cite{taxcross} specified three supervised tasks, mapping between one base ``site'' (Prisma 30) and the three target ``sites'' (Prisma 60, Connectom 30, and Connectom 60). We modify this task, removing correspondence/pairing knowledge between sites (keeping this information for validation and testing), and including the inverse mapping task (target to base).
This results in six tasks, two for each target site.

We train a ``single-site'' network for each of the six tasks, learning representations for Prisma 30 and a single target site, a multi-site variant across all six tasks. \NewForRevision{During training the method is not provided corresponding patches, and is \emph{only} given individual patches. A single sample corresponds to one patch, not a pair of patches. Paired patches are only used to calculate error measures.}

We measure the performance of each method on the holdout set of subjects using the Root Mean Squared Error (RMSE) between each method's output and the ground truth target images in the original DWI basis (after pre-processing). For comparison we also include results from Mirzaalian et al. \cite{mirzaalian2018multi}, which is the only other unsupervised method we are aware of in the literature.

We further assess the performance of each method by estimating the fiber orientation distributions (FODs) for each reconstruction using Multi-Shell Multi-Tissue Constrained Spherical Deconvolution (MSMT-CSD) \cite{jeurissen2014multi}, with response functions estimated using the method proposed in Dhollander et al. \cite{dhollander2016unsupervised}. For both of these steps we use the implementations from MRtrix3 \cite{tournier2019mrtrix3}. For each FOD we compute the maxima at each voxel and compare it to the closest maxima of the ground truth image to compute angular error.

\NewForRevision{In order to assess the fidelity of common local diffusion model summary measures before and after harmonization, we measure the Mean Average Percent Error (Mean APE) and the Coefficient of Variation (CV) between method-estimated and observed summary measures, reported in Table \ref{tab:APE}. We measured Mean APE and CV for Fractional Anisotropy, Mean Diffusivity, Mean Kurtosis \cite{jensen2005diffusional} , and Return-to-Origin-Probability (RTOP) \cite{ozarslan2013mean}. This mirrors the analysis in Ning et al. 2019 \cite{ning2019cross}.}

In order to test the specific effects of our compressive regularizations, we conducted two ablation tests of our method, comparing it to the ``regular networks'' with parameters described in Section \ref{subsec:config}. We re-trained both single-site and multi-site methods with the invariance parameter $\lambda$ set to 0, but otherwise the same settings. We further trained two more networks for $\lambda$ and $\alpha$ set to 0. Effectively this ablates the added invariance-inducing compressive elements of the loss function. We then compared their performance by computing the voxelwise difference in RMSE in each heldout test subject.

For the proposed methods and corresponding ablated networks we assessed the amount to which we removed site information from of the learned representation $z$ by attempting predict $s$ from $z$. If there is no information in $z$ about $s$ then we would expect the optimal predictor to do no better than random. 
To this end we trained feed forward networks to predict $s$ from $z$ (``post-hoc adversaries''). As shown in Moyer et al. \cite{NIPS2018_8122}, the cross-entropy error of these networks is a lower bound for the mutual information $I(s,z)$. The post-hoc adversaries had same configuration as the patch-adversaries (two $32$-unit layers using $\text{tanh}(\cdot)$ activations and the softmax cross-entropy loss).

\subsection{Configuration and Parameters}
\label{subsec:config}

We implemented our method for image patches composed of a center voxel and each of its six  immediate neighbors. Each of these voxels has two shells of DWI signal, which we mapped to the SH $8^{th}$ order basis, plus one $b_0$ channel. Unravelling these patches and shells, the input is then a vector with $91\times 7 = 637$ elements.

We use three-layer fully connected neural networks for encoder $q(z|x)$ and conditional decoder $p(x|z,s)$, with $256$, $128$, $64$ hidden units respectively for the encoder, and the reverse ($64$, $128$, then $256$) for the decoder. The latent code $z$ is parameterized by a $32$ unit Gaussian layer ($z$). This layer is then concatenated with the scanner/protocol one-hot representation $s$, and input into the decoder. We use $\text{tanh}(x)$ transformations at each hidden layer, with sigmoid output from the encoder for the variance of the Gaussian layer. The adversary is a fully connected two-layer network with 32 units at each layer, with $\text{tanh}(x)$ units again at each hidden node.

For each task we train our network for 1000 epochs, which took 19 hours to train in the pair-wise case on standard desktop equipped with an external Nvidia Titan-Xp with 12GB of RAM using TensorFlow (32GB of CPU RAM, 4 cores). We loosely tune the hyper parameters so losses are approximately on the same order of magnitude, with $\alpha=1.0$, $\beta=1.0$, $\gamma=10.0$, and $\lambda=0.01$. We use these same parameters for both the pair-wise tasks as well as the multi-task experiments. We use an Adam learning rate of $0.0001$ and a batch size of 128. For each batch provided for primary network training we provide 10 epochs for  training the adversary.

\section{Results}

\label{sec:results}

\begin{figure}[ht!]
\centering
\includegraphics[width=0.85\textwidth]{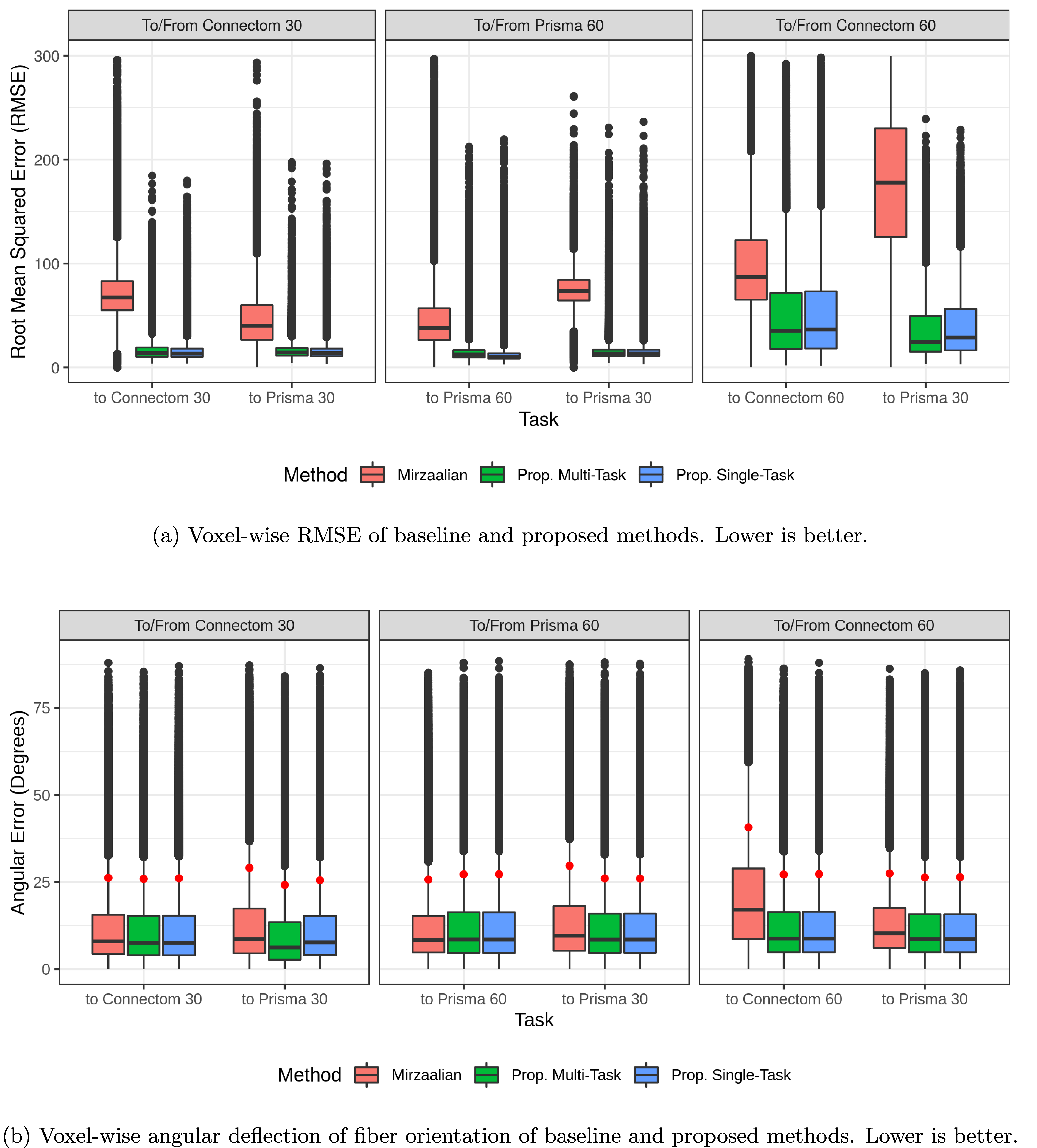}
\caption{\CapOpt{ \NewForRevision{Here we plot the \textbf{voxel-wise} performance as measured by RMSE (\textbf{top}) and angular deflection (\textbf{bottom}), i.e. the angular difference between the global maxima of the Fiber Orientation Distributions recovered by MSMT-CSD \cite{jeurissen2014multi} from the ground-truth and reconstructed images, measured in degrees. This is shown for the Mirzaalian et al. \cite{mirzaalian2018multi} method as well as our two proposed methods on each of the six harmonization tasks (Prisma 30 to and from each of the other scanner/site combinations). All RMSE values are calculated in the original signal representation. In both plots \textbf{lower} is better. For the angular deflection error, the $90^{th}$ percentile data point plotted in {\color{red} red}.} }}
\refstepcounter{subfigure}\label{fig:AtoAll}
\refstepcounter{subfigure}\label{fig:angles}
\label{fig:2super}
\end{figure}

\begin{table}[ht]
\centering
\begin{tabular}{| c | c | l | c | c | c | c | c | c | c | c | }
\cline{4-11}
\multicolumn{3}{c | }{} & \multicolumn{2}{c | }{FA} & \multicolumn{2}{ c |}{MD} & \multicolumn{2}{ c |}{MK} & \multicolumn{2}{ c |}{RTOP} \\\hline
* & P30 & Method & APE & CV & APE & CV & APE & CV & APE & CV \\ 
  \hline
\multirow{6}{*}{\rotatebox[origin=c]{90}{Connectom 30}}
& \multirow{3}{*}{\rotatebox[origin=c]{90}{to}}
& Mirzaalian & 0.46 & 0.48 & 0.42 & 0.80 & 0.99 & 1.02 & 0.19 & 1.22 \\ 
& & Single-task & 0.25 & 0.26 & 0.12 & 0.17 & 3.37 & 0.30 & 0.11 & 0.28 \\ 
& & Multi-task & 0.28 & 0.30 & 0.12 & 0.16 & 3.15 & 0.28 & 0.11 & 0.16 \\ 
\cline{2-11}
& \multirow{3}{*}{\rotatebox[origin=c]{90}{from}}
& Mirzaalian & 0.50 & 0.39 & 0.52 & 0.59 & 0.96 & 1.05 & 0.22 & 0.26 \\ 
& & Single-task & 0.29 & 0.24 & 0.22 & 0.18 & 3.72 & 0.30 & 0.13 & 0.18 \\ 
& & Multi-task & 0.30 & 0.27 & 0.21 & 0.17 & 3.90 & 0.31 & 0.12 & 0.17 \\ 
\hline
\multirow{6}{*}{\rotatebox[origin=c]{90}{Prisma 60}}
& \multirow{3}{*}{\rotatebox[origin=c]{90}{to}}
& Mirzaalian & 0.52 & 0.55 & 0.60 & 0.67 & 0.99 & 1.05 & 0.29 & 0.41 \\ 
& & Single-task & 0.34 & 0.37 & 0.12 & 0.23 & 3.44 & 0.31 & 0.14 & 1.47 \\
& & Multi-task & 0.34 & 0.37 & 0.12 & 0.19 & 3.17 & 0.29 & 0.13 & 0.45 \\ 
\cline{2-11}
& \multirow{3}{*}{\rotatebox[origin=c]{90}{from}}
& Mirzaalian & 0.64 & 0.45 & 0.48 & 0.44 & 0.96 & 0.98 & 0.22 & 0.28 \\ 
& & Single-task & 0.41 & 0.30 & 0.38 & 0.15 & 0.48 & 0.14 & 0.09 & 0.16 \\ 
& & Multi-task & 0.42 & 0.32 & 0.38 & 0.15 & 0.45 & 0.13 & 0.08 & 0.11 \\ 
\hline
\multirow{6}{*}{\rotatebox[origin=c]{90}{Connectom 60}}
& \multirow{3}{*}{\rotatebox[origin=c]{90}{to}}
& Mirzaalian & 0.86 & 0.79 & 0.88 & 0.92 & 1.01 & 1.05 & 1.11 & 26.18 \\ 
& & Single-task & 0.35 & 0.36 & 0.26 & 0.81 & 3.22 & 0.35 & 0.16 & 0.50 \\ 
& & Multi-task & 0.35 & 0.36 & 0.14 & 0.23 & 3.31 & 0.36 & 0.14 & 0.37 \\ 
\cline{2-11}
& \multirow{3}{*}{\rotatebox[origin=c]{90}{from}}
& Mirzaalian & 3.19 & 1.26 & 4.64 & 5.97 & 0.88 & 0.90 & 0.63 & 0.69 \\ 
& & Single-task & 1.86 & 0.40 & 2.93 & 0.31 & 4.44 & 0.27 & 0.10 & 0.17 \\ 
& & Multi-task & 1.77 & 0.38 & 2.84 & 0.29 & 4.56 & 0.27 & 0.10 & 0.32 \\ 
\hline
\end{tabular}
\caption{\CapOpt{\NewForRevision{Here we report the mean Absolute Percent Error (APE) and mean Coefficient of Variation (CV) per voxel for each of the methods for four common diffusion summary measures: Fractional Anisotropy (FA), Mean Diffusivity (MD), Mean Kurtosis (MK) \cite{jensen2005diffusional} , and Return-to-Origin-Probability (RTOP) \cite{ozarslan2013mean}. The APE measure is the same as the error metric reported in Ning et al. \cite{ning2019cross}; similar to Ning et al. \cite{ning2019cross}, we report values as decimals (where $1.00$ corresponds to 100\%), and not actual percentages. It is well known that the APE measure is biased towards methods reporting smaller values \cite{armstrong1992error,makridakis1993accuracy}. We therefore also report the Coefficient of Variation, computed by dividing the RMSE by the observed sample mean. This measure is also sometimes referred to as the Relative RMSE. }}}
\label{tab:APE}
\end{table}

\begin{table}
\centering
\begin{tabular}{ | c | c | l | c | c | c | c | }
  \hline
* & P30 & Method & FA PE & MD PE & MK PE & RTOP PE \\ 
  \hline
\multirow{6}{*}{\rotatebox[origin=c]{90}{Connectom 30}}
& \multirow{3}{*}{\rotatebox[origin=c]{90}{to}}
& Mirzaalian  & -0.27 & -0.26 & -0.87 & -0.09 \\ 
& & Single-task & -0.05 & 0.02 & 3.31 & -0.01 \\ 
& & Multi-task & -0.11 & 0.03 & 3.04 & -0.03 \\ 
\cline{2-7}
& \multirow{3}{*}{\rotatebox[origin=c]{90}{from}}
& Mirzaalian & 0.22 & -0.45 & -0.95 & 0.20 \\ 
& & Single-task & 0.10 & 0.18 & 3.57 & -0.07 \\ 
& & Multi-task & 0.04 & 0.15 & 3.79 & -0.04 \\ \hline 
\multirow{6}{*}{\rotatebox[origin=c]{90}{Prisma 60}}
& \multirow{3}{*}{\rotatebox[origin=c]{90}{to}}
& Mirzaalian & -0.15 & -0.58 & -0.93 & -0.19 \\ 
& & Single-task & -0.15 & -0.00 & 3.38 & 0.00 \\
& & Multi-task & -0.14 & 0.05 & 3.03 & -0.05 \\ 
\cline{2-7}
& \multirow{3}{*}{\rotatebox[origin=c]{90}{from}}
& Mirzaalian & 0.48 & -0.22 & -0.94 & 0.20 \\ 
& & Single-task & 0.18 & 0.29 & 0.44 & 0.03 \\
& & Multi-task & 0.15 & 0.32 & 0.33 & -0.01 \\ \hline
\multirow{6}{*}{\rotatebox[origin=c]{90}{Connectom 60}}
& \multirow{3}{*}{\rotatebox[origin=c]{90}{to}}
& Mirzaalian & 0.23 & -0.88 & -0.91 & 0.36 \\ 
& & Single-task & 0.02 & 0.16 & 3.06 & -0.04 \\
& & Multi-task & -0.02 & 0.01 & 3.16 & -0.00 \\  
\cline{2-7}
& \multirow{3}{*}{\rotatebox[origin=c]{90}{from}}
& Mirzaalian & 2.32 & 3.65 & -0.73 & -0.62 \\ 
& & Single-task & 1.70 & 2.88 & 4.33 & -0.02 \\ 
& & Multi-task & 1.56 & 2.77 & 4.48 & 0.01 \\ \hline
\multicolumn{2}{c | }{} & \multicolumn{5}{ c | }{ Negative PE implies Actual Value $>$ Estimated Value } \\
\cline{3-7}
\end{tabular}
\caption{\CapOpt{\NewForRevision{Here we report the mean Percent Error (PE) per voxel for each of the methods for four common diffusion summary measures. Negative PE implies that the value from the real data was greater than the value from the harmonization method.}}}
\label{tab:exp}
\end{table}

\begin{figure}[h]
    \begin{subfigure}[b]{0.5\textwidth}
        \centering
        \includegraphics[height=7cm]{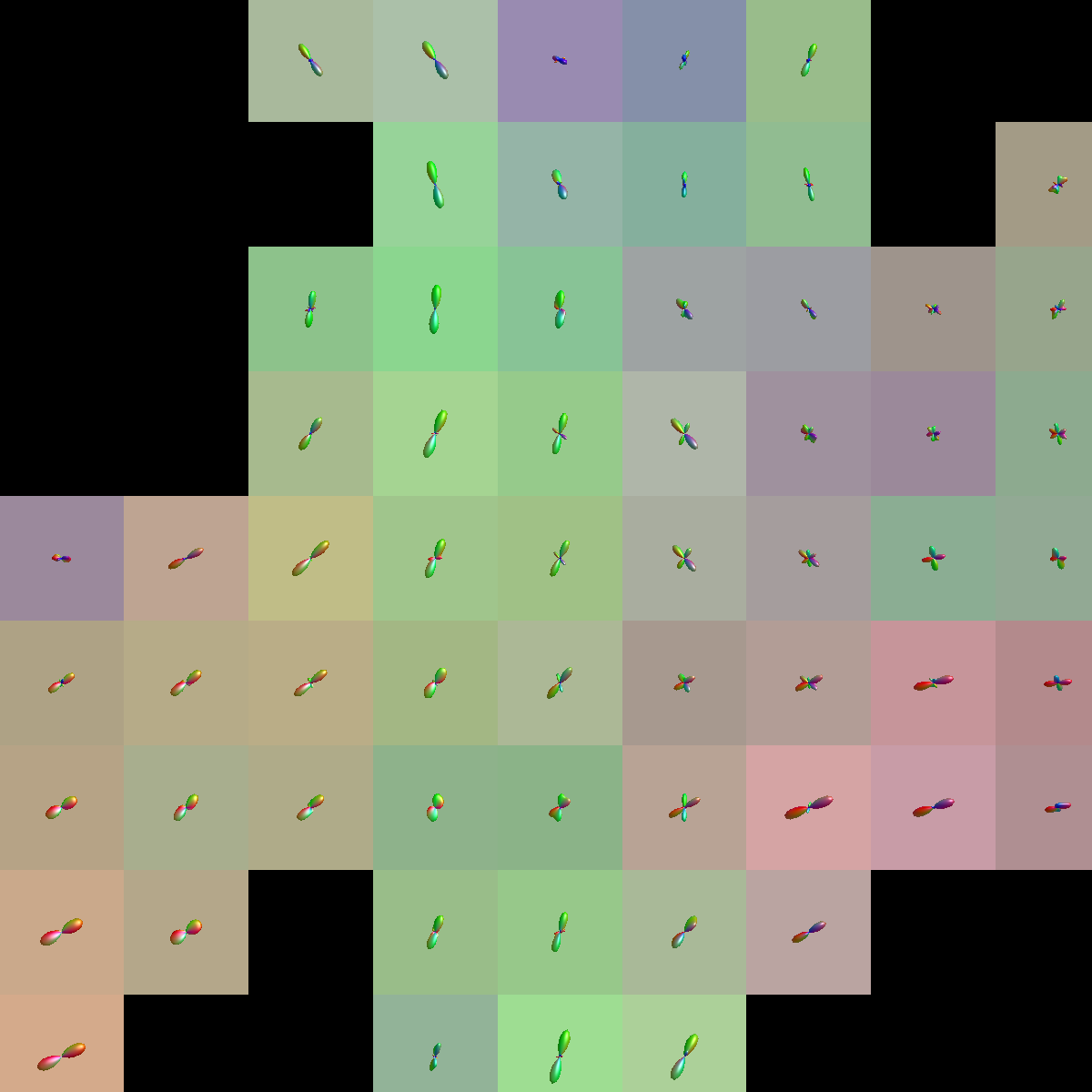}
        \caption{Target image (Prisma 30).}
    \end{subfigure}
    \begin{subfigure}[b]{0.5\textwidth}
        \centering
        \includegraphics[height=7cm]{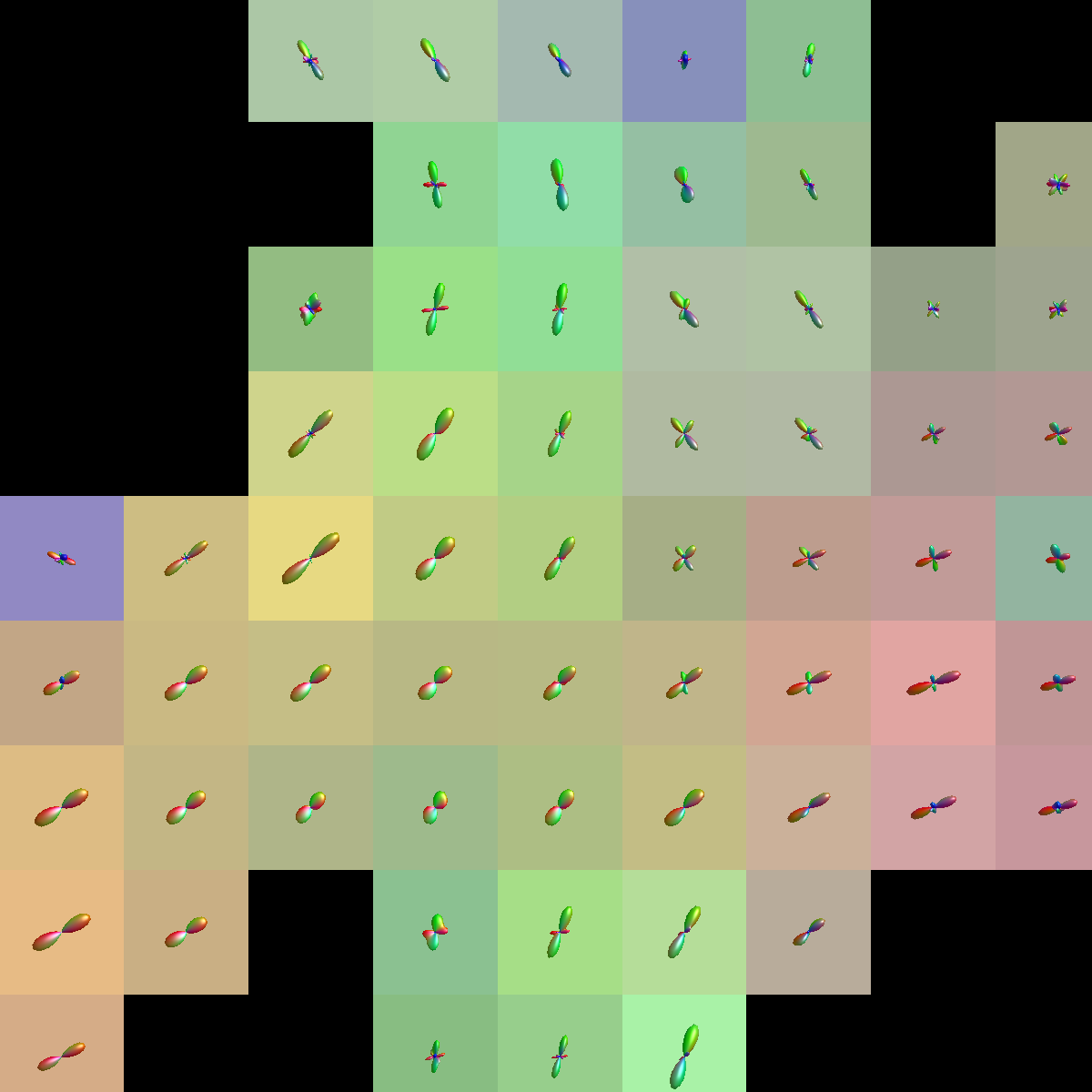}
        \caption{Reconstruction using Mirzaalian et al. method.}
    \end{subfigure}
    \begin{subfigure}[b]{0.5\textwidth}
        \centering
        \includegraphics[height=7cm]{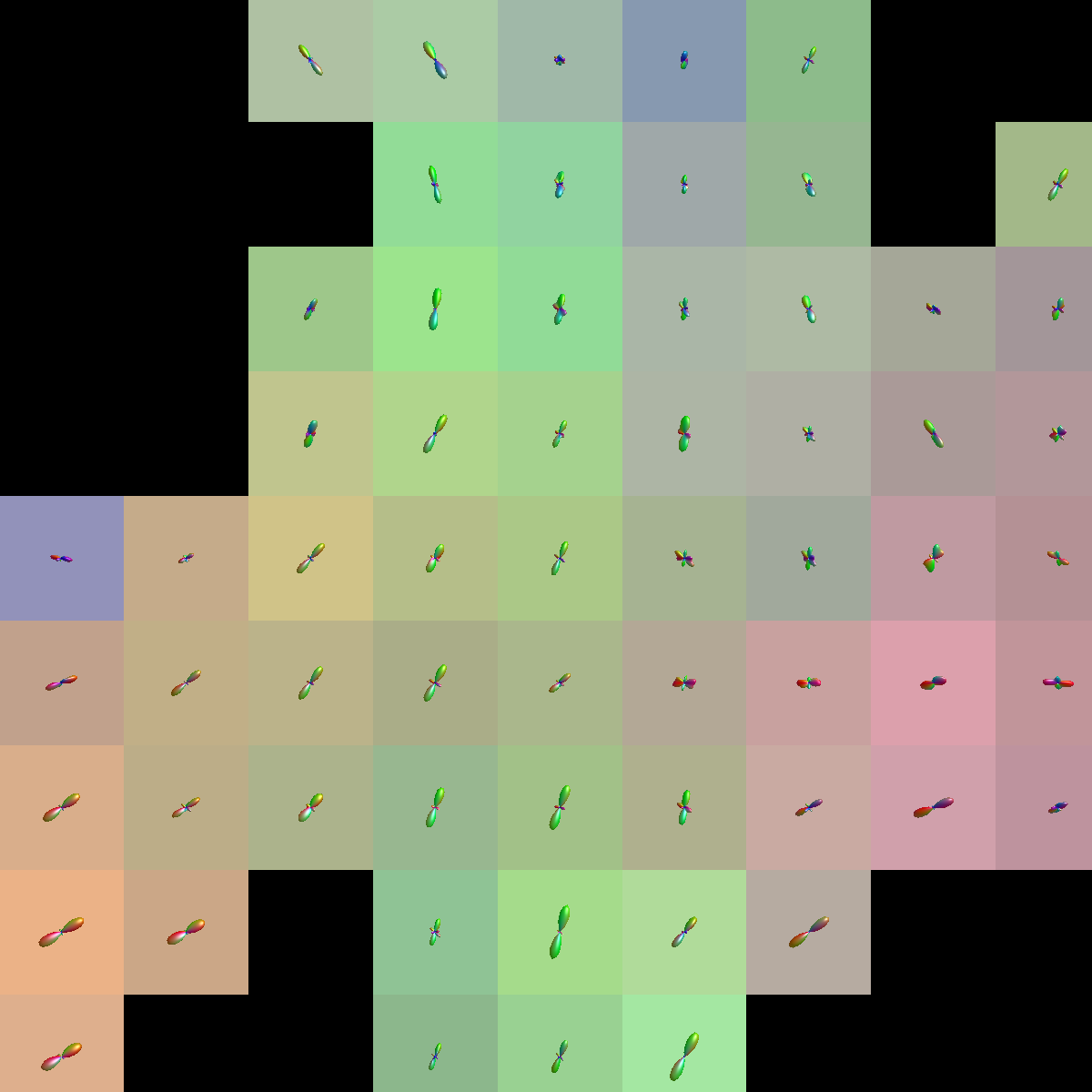}
        \caption{Reconstruction using single-site method.}
    \end{subfigure}
    \begin{subfigure}[b]{0.5\textwidth}
        \centering
        \includegraphics[height=7cm]{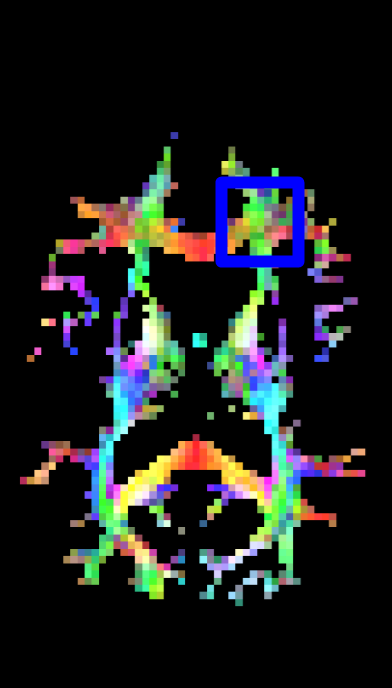}
        \caption{Whole brain, subset location in {\color{blue} blue}.}
    \end{subfigure}
    \caption{\CapOpt{Here we plot exemplar FOD glyphs (estimated via MSMT-CSD) from the actual data, a reconstruction using the Mirzaalian et al. method, and a reconstruction using the proposed single-site method. Inputs to each reconstruction were the data from the Prisma 60 protocol/site. The background colors represent the direction of the FOD maxima.}}
    \label{fig:glyphs}
\end{figure}

Figure \ref{fig:AtoAll} plots the root mean squared error (RMSE) by voxel of the baseline, single-site proposed method, and multi-site proposed method, as evaluated on the holdout test subjects, \NewForRevision{ in the original signal representation}. Our proposed methods show improvement over the baseline method in each case. In the pair-wise task between similar protocols (mapping between Prisma 30 and Connectom 30), these improvements have non-overlapping inner quartile range. For dissimilar protocols, i.e. mapping between Prisma 30 and Prisma 60 or Connectom 60, our proposed method shows improvements, though the difference is less pronounced. Surprisingly, for higher resolution target images the multi-site method performs as well or better than the pair-wise method and the baseline; this may be due to the multi-task method receiving many more volumes overall, allowing it to gather more information (albeit biased by other scanners) or preventing it from overfitting.

Figure \ref{fig:angles} plots the voxel-wise angular deflection of each method, as measured by MSMT-CSD. For Connectom 30 and Prisma 60, both to and from Prisma 30, all three methods are comparable, with median errors well below $20^\circ$, and $90^{th}$ percentile errors all slightly above $25^\circ$. For mappings to the Connectom 60 protocol, the Mirzaalian et al. method has generally higher error, though the inner quartile ranges still overlap for all methods. We plot a subset of the FODs from the original image and two of the reconstructions in Figure \ref{fig:glyphs}.

\begin{figure}[t!]
    \centering
    \includegraphics[width=0.8\textwidth]{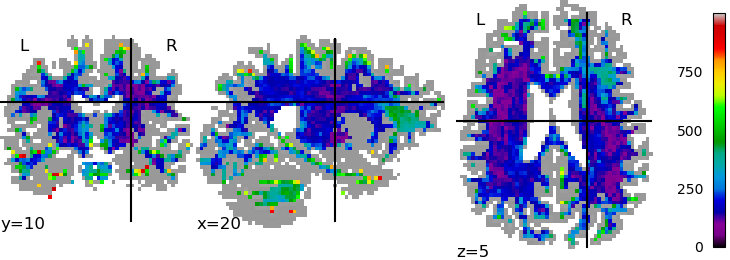}
    \includegraphics[width=0.8\textwidth]{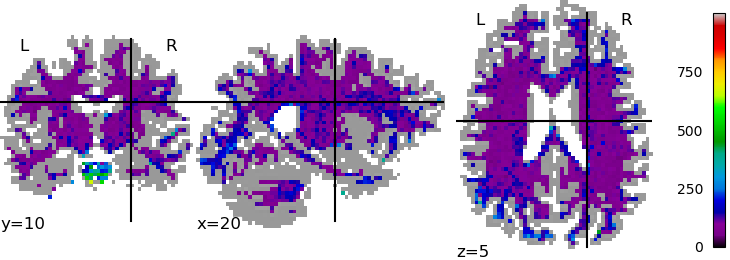}
    \includegraphics[width=0.8\textwidth]{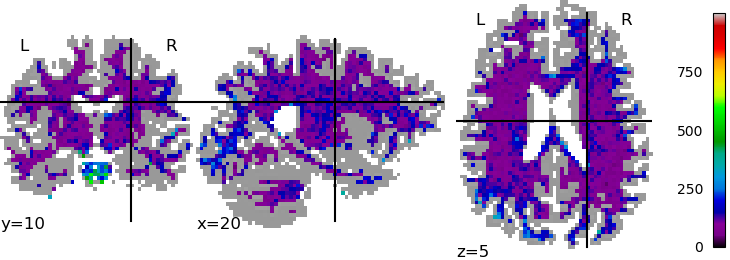}
    \caption{\CapOpt{We plot the spatial distribution of RMSE per voxel, displayed in slices centered at (x,y,z) = (10,22,35) for mappings from the Prisma 30 protocol to the Connectom 30 protocol, for (\textbf{top row}) the Mirzaalian \cite{mirzaalian2018multi} baseline, (\textbf{center row}) the single-site proposed method, and (\textbf{bottom row}) the multi-site proposed method. The color scale is the same between the rows, as well as between this figure, Fig. \ref{fig:rmseAB}, and Fig. \ref{fig:rmseAD}. \NewForRevision{All RMSE values are calculated in the original signal representation.}} }
    \label{fig:rmseAC}
\end{figure}

\begin{figure}[t!]
    \centering
    \includegraphics[width=0.8\textwidth]{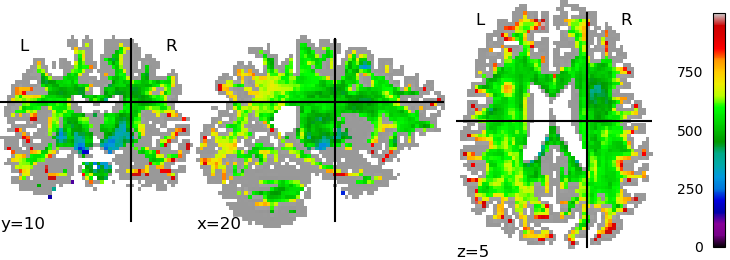}
    \includegraphics[width=0.8\textwidth]{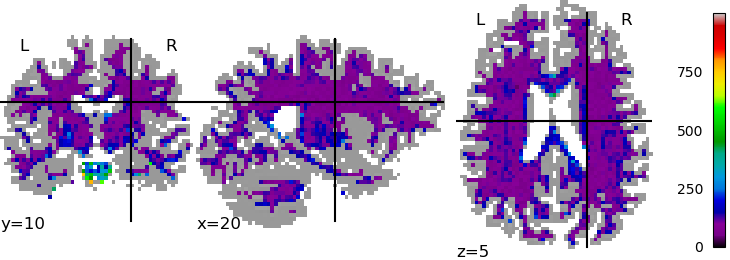}
    \includegraphics[width=0.8\textwidth]{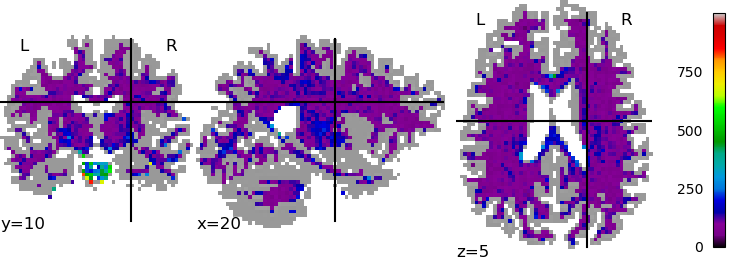}
    \caption{\CapOpt{We plot the spatial distribution of RMSE per voxel, displayed in slices centered at (x,y,z) = (10,22,35) for mappings from the Prisma 30 protocol to the Prisma 60 protocol, for (\textbf{top row}) the Mirzaalian \cite{mirzaalian2018multi} baseline, (\textbf{center row}) the single-site proposed method, and (\textbf{bottom row}) the multi-site proposed method. The color scale is the same between the rows, as well as between this figure, Fig. \ref{fig:rmseAC}, and Fig. \ref{fig:rmseAD}. \NewForRevision{All RMSE values are calculated in the original signal representation.} }}
    \label{fig:rmseAB}
\end{figure}

\begin{figure}[t!]
    \centering
    \includegraphics[width=0.8\textwidth]{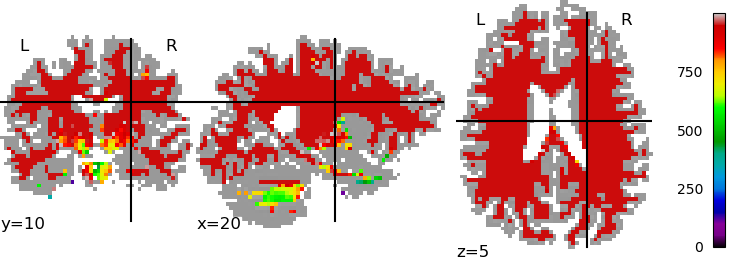}
    \includegraphics[width=0.8\textwidth]{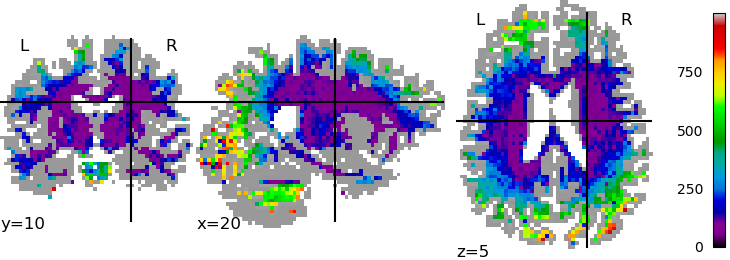}
    \includegraphics[width=0.8\textwidth]{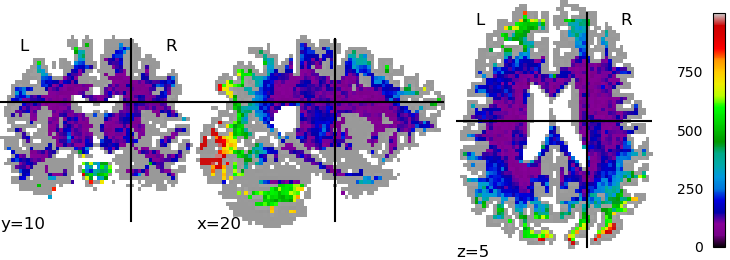}
    \caption{\CapOpt{We plot the spatial distribution of RMSE per voxel, displayed in slices centered at (x,y,z) = (10,22,35) for mappings from the Prisma 30 protocol to the Connectom 60 protocol, for (\textbf{top row}) the Mirzaalian \cite{mirzaalian2018multi} baseline, (\textbf{center row}) the single-site proposed method, and (\textbf{bottom row}) the multi-site proposed method. The color scale is the same between the rows, as well as between this figure, Fig. \ref{fig:rmseAC}, and Fig. \ref{fig:rmseAB}. \NewForRevision{All RMSE values are calculated in the original signal representation.} }}
    \label{fig:rmseAD}
\end{figure}

Figures \ref{fig:rmseAC}, \ref{fig:rmseAB}, and \ref{fig:rmseAD} show the spatial distribution of the error for each tested method on a single test subject, for mappings between Prisma 30 and Connectom 30, Prisma 60, and Connectom 60 respectively. For the Prisma 30 to Connectom 30 mapping, overall the Mirzaalian baseline \cite{mirzaalian2018multi} has higher error than the other methods as shown by the overall coloring. The Mirzaalian baseline \cite{mirzaalian2018multi} and the multi-site proposed method show significant white matter patterning (though in varying degree); optimally we would like to see uncorrelated residuals, like those shown in the single-site method. 

The Connectom 60 error plots (Fig. \ref{fig:rmseAD}) have a strong spatial patterns at both the occipital and frontal poles, shown in all methods. This wide-scale effect is somewhat mitigated by the proposed methods, but is still present in all error distributions.

\NewForRevision{Table \ref{tab:APE} reports the Absolute Percent Error (APE) and the estimated Coefficient of Variation (CV) for each method voxel-wise for four commonly used diffusion summary measures: Fractional Anisotropy (FA), Mean Diffusivity (MD), Mean Kurtosis (MK) \cite{jensen2005diffusional}, and Return-to-Origin-Probability (RTOP) \cite{ozarslan2013mean}. It is well known that APE is biased towards methods reporting smaller values, and becomes inaccurate and inflated as actual observed values approach zero \cite{armstrong1992error,makridakis1993accuracy}. In our context this means that for FA and MD, more spherical tensors are weighted strongly, while more anisotropic tensors are weighted less. Due to this bias, we also report the estimated Coefficient of Variation (CV) \cite{abdi2010coefficient}, which is the RMSE divided by the observed sample mean\footnote{For unbiased estimates the RMSE divided by sample mean should further be multiplied by a factor of $(\sqrt{\frac{N}{N-1}})(1 - \frac{1}{4N})$, where $N$ is the number of tested voxels. However, this number is very close to $1$, and the resulting change is negligible.}. CV has also been used to assess summary statistic variation between scanners \cite{cercignani2003inter,vollmar2010identical}, and is sometimes referred to as Relative RMSE. }

\NewForRevision{ For all reported summary measures except MK, the proposed methods map to and from Connectom 30 perform well under both error measures. Mapping to both Connectom 60 and Prisma 60 from Prisma 30 has higher error than the converse (Prisma 30 to Connectom/Prisma 60); this fits our intuitions about upsampling, as both ``60'' protocols have higher angular resolution.}

\NewForRevision{For Mean Kurtosis in remapped scans to/from Connectom 30, the APE is very high while the CV is surprisingly low. This pattern is also seen in FA, MD, and MK for scans mapped to Connectom 60, and for MK in scans mapped from Prisma 60. Because the APE error is above 100\% (but CV is small), we believe that the methods are overestimating small actual values, since underestimation error is bounded at 100\% for non-negative measures. In order to further verify this, we computed the Percent Error (without absolute values) shown in Table \ref{tab:exp}, indicating the average bias above or below the actual observed value. Since the MK PE for both proposed methods is very close to the APE, this indicates that on average small values are being overestimated. Discussion continues in Section \ref{sec:disc}.}

\subsection{Ablation Test Results and Post-hoc Adversarial Accuracies}
\label{subsec:ablation}

\begin{table}[h]
\begin{center}
\begin{tabular}{ | c | p{1cm}|p{1cm} | p{1cm}|p{1cm} | p{1cm}|p{1cm} |}
\multicolumn{1}{c}{} & \multicolumn{6}{ c }{ $\Delta \text{RMSE}$ for $\lambda=0$}\\ \cline{2-7}
\multicolumn{1}{c |}{} & \multicolumn{6}{ c | }{ $\Delta \text{RMSE} = \text{RMSE}_{\text{reg.}} - \text{RMSE}_{\lambda=0}$ } \\\cline{2-7}
\multicolumn{1}{c |}{} & \multicolumn{2}{ c | }{Connectom 30} & \multicolumn{2}{ c | }{Prisma 60}  & \multicolumn{2}{ c | }{Connectom 60}\\ \cline{2-7}
\multicolumn{1}{c|}{} & To & From & To & From & To & From \\\hline
Proposed Single-task 
& 1.2 & 2.6 & -1.4 & 7.6 & -4.0 & -46.8 
\\ \hline
Proposed Multi-task 
 & 6.3 & 2.6 & 16.7 & 12.25 & -4.1 & -65.7\\ \hline \multicolumn{7}{ c }{} \\
\multicolumn{1}{c}{} & \multicolumn{6}{ c }{ $\Delta \text{RMSE}$ for $\lambda=0,\alpha=0$}\\ \cline{2-7}
\multicolumn{1}{c|}{} & \multicolumn{6}{ c | }{ $\Delta \text{RMSE} = \text{RMSE}_{\text{reg.}} - \text{RMSE}_{\lambda=0,\alpha=0}$ } \\\cline{2-7}
\multicolumn{1}{c|}{} & \multicolumn{2}{ c | }{Connectom 30} & \multicolumn{2}{ c | }{Prisma 60}  & \multicolumn{2}{ c | }{Connectom 60}\\ \cline{2-7}
\multicolumn{1}{c|}{} & To & From & To & From & To & From \\\hline
Proposed Single-task 
& 6.1 & 1.9 & -6.4 & -1.5 & -17.0 & -55.7 
\\ \hline
Proposed Multi-task 
 & 94.3 & 89.4 & -324.8 & -354.8 & -217.0 & -90.0\\ \hline
\end{tabular}
\end{center}
\caption{\CapOpt{Here we report the mean per-voxel test set RMSE change between the regular model and two ablated models, where (\textbf{top}) the invariance term $\lambda$ was set to zero, and (\textbf{bottom}) the invariance term $\lambda$ and the VAE prior term $\alpha$ were set to zero. Negative values indicate that the regular model has better performance than the ablated models.}}
\label{tab:ablation}
\end{table}

\begin{table}[h]
\begin{center}
\begin{tabular}{ | c | p{2cm} | p{2cm} | p{2cm} | p{2cm} |}
\multicolumn{1}{c}{} & \multicolumn{4}{ c }{ Post-hoc Adversarial Accuracy, Predicting $s$ from $z$ }\\ \cline{2-5}
\multicolumn{1}{c |}{} & Best Possible & Full Model &  $\lambda = 0$  & $\lambda = 0, \alpha = 0$\\ \hline
Proposed Single-task, Connectom 30 
& 0.5 & 0.61 & 0.63 & 0.63
\\ \hline
Proposed Single-task, Prisma 60 
& 0.5 & 0.5 & 0.51 & 0.54
\\ \hline
Proposed Single-task, Connectom 60 
& 0.5 & 0.63 & 0.68 & 0.85
\\ \hline
Proposed Multi-task 
& 0.25  & 0.41 & 0.41 & 0.62 \\ \hline
\end{tabular}
\end{center}
\caption{\CapOpt{Here we report the mean per-patch test set classification accuracy for an adversary trained post-hoc to predict the site variable $s$ from the latent representation $z$, for $z$ taken from the full model (\textbf{center left column}), the $\lambda$ ablated model (\textbf{center right column}), and the $\lambda$ and $\alpha$ ablated model (\textbf{right column}). The \textbf{far left column} shows the best possible performance. The architecture and training protocol for the adversaries is described in Section \ref{subsec:protocol}. Here \textbf{lower} is better (``closer to invariant'').}}
\label{tab:post-hoc-adv}
\end{table}

Table \ref{tab:ablation} shows the results of the ablation tests, where we set either $\lambda$ to zero or both $\lambda$ and $\alpha$ to zero, effectively removing invariance and prior terms respectively from the primary loss function. We computed the difference in the RMSE between the regular model and the ablated models on the hold-out test dataset. For Connectom 30 both to and from Prisma 30, both the invariance term and the prior term hinder reconstruction performance. When only the invariance term is removed this effect is slight, but when both are removed the effect is much stronger in the multi-task setting.

For Prisma 60 mappings differences without the invariance term for the single task method are relatively small, while removing both the invariance and prior terms leads to large increases in RMSE. For mappings to Connectom 60, differences in RMSE follow a similar pattern to Prisma 60, with performance decreasing with both invariance and prior terms ablated. For the mappings from Connectom 60, performance strongly drops without the invariance term and then further drops without the prior term.

Table \ref{tab:post-hoc-adv} shows the results of the post-hoc adversarial predictions. Setting $\lambda=0$ uniformly increases post-hoc adversarial accuracy, and setting $\alpha=0$ increases accuracy further in both Prisma 60 and Connectom 60 cases. For the multi-site model the prediction task is considerably harder, yet setting both $\lambda$ and $\alpha$ to zero induces relatively high adversarial accuracy in the multi-task setting ($\sim 60$\%).

It is unsurprising that the invariance term does not aid in reconstruction for more similar protocols/scanners. Inclusion of this term should lead to more compression, and thus less information in $z$ relevant to $x$, which in turn should lead to worse reconstruction. Further, this intuition also extends to the VAE prior term, which is a sufficient condition for the compressive portion of the invariance term (if $\mathcal{L}_{prior} = 0$ then $KL[q(z|x)\|q(z)] = 0$). It is interesting, however, that these terms lead to \emph{increased} performance for dissimilar protocols/scanners i.e. Connectom 60 and Prisma 60. This indicates that these two loss terms are helpful for generalization.


\section{Discussion}
\label{sec:disc}

\NewForRevision{Our proposed harmonization method is unsupervised in that we \emph{do not} require multiple images from the same subject or phantom from separate sites (i.e. paired data) in order to train our method. It is advisable to validate using such data, but due to the expense of collecting images from the same subject at varying sites it is advantageous to limit reliance on these data.}

We believe it is important to understand the trade-off between reconstructive error and adversarial accuracy (e.g. between performance in Figure \ref{fig:AtoAll} and \ref{tab:post-hoc-adv}). It is obviously desirable to have high reconstructive accuracy, yet any attempt to induce invariance necessarily removes information (i.e. site information), which reduces this accuracy. At the other end of the spectrum, there is always a family of perfectly invariant solutions (constant images), but these also have no information about the subjects, and subsequently very high reconstructive error. It is thus important to consider both in selecting a remapping method.

Because of the VAE prior's sufficiency for compressing $z$, empirically we can create an acceptable method without the invariance term (i.e. with $\lambda=0$). This agrees with our intuition about Eq. \ref{eq:IZS}, where compression plus conditional reconstruction is a proxy for invariance. It appears that the exact form of compression is less impactful. However, best performance is achieved by including an invariance term.


It is tempting to attempt to interpret the encodings $z$, but these efforts should not be undertaken lightly. The encoding and decoding functions are designed to be non-linear, and individual components of $z$ may have interaction effects with other components. Further, the encodings $z$ are not images or patches, lacking a spatial domain. With careful construction analysis may be possible, but it is almost certainly non-trivial to do in the encoding domain.

In the current method we reconstruct images for a specific target site $s'$. We might instead look for a site agnostic image. This is philosophically challenging: images are by nature collected at sites, and there are no site-less images. While we can manipulate our method to produce an $s^*$ average site, the output image may not be representative of any of the images. It may be that all images must have site information, and that the quotient representation is not an image at all. On the other hand, for other tasks $y$ that are not images, e.g., prediction of pathology or prognosis, we can use $z$ to make unbiased (scanner-agnostic) predictions of $y$. In cases where the actual goal is not in the image domain (for which the harmonization task is a pre-processing step), such a formulation may be beneficial, and could be built from our proposed method.

\subsection{Limitations}

This method cannot remove long-range scanner-biases; this is due to the patch-based architecture.
In theory, with larger patches, we could avoid this limitation; current hardware, in particular GPU memory and bus speeds, limit our computation to small patches for dMRI. Specific work in this domain has been done to reduce memory load \cite{blumberg2018deeper}, but it is by no means solved, especially for high angular resolution data such as the HCP dataset  \cite{sotiropoulos2013advances}. We hypothesize that a similar architecture with larger patches or whole images could rectify this particular problem - architectures that may become accessible with increased hardware capabilities - or better model compression/computational reduction techniques.

\NewForRevision{In the present work, the proposed method was only evaluated on white matter, and not in grey matter (neither cortex nor subcortical structures). White matter analyses generally focus on models of restricted axonal compartments (fibers), with derived measures such as fiber orientation distribution functions (FODs) and voxel-wise data with generally high anisotropy. Grey matter analyses in contrast may focus more on signal from isotropic compartments and/or dendritic arbors \cite{jespersen2007modeling}, and notably their models may be robust or vulnerable to site-bias in different ways. We have not considered grey matter signal or model summary statistics in this analysis, and thus we advise caution when applying this method to identified grey matter voxels. 
Further, as Table \ref{tab:APE} and Table \ref{tab:exp} show, for low values of Mean Kurtosis the proposed method is inaccurate and has a positive bias in reconstruction. We advise caution when using this method where the accuracy of these measures for low relative values is critical.}



\section{Conclusion}

In the present work we have constructed a method for learning scanner-invariant representations. These representations can then be used to reconstruct images under a variety of different scanner conditions, and due to the data processing inequality the reconstruction's mutual information with the original scanner will be low. This we demonstrate to be useful for the unsupervised case of data harmonization in diffusion MRI; critically, we can harmonize data without explicit pairing between images, reducing the need for. Surprisingly in some cases the multi-task method outperforms a pairwise method with similar architecture. This may hint at further benefits for learning shared representations.

\subsection*{Acknowledgements}

This work was supported by NIH grants P41 EB015922, R01 MH116147, R56 AG058854, RF1 AG041915, and U54 EB020403, DARPA grant W911NF-16-1-0575, as well as the NSF Graduate Research Fellowship Program Grant Number DGE-1418060, and a GPU grant from NVidia.

The data were acquired at the UK National Facility for In Vivo MR Imaging of Human Tissue Microstructure located in CUBRIC funded by the EPSRC (grant EP/M029778/1), and The Wolfson Foundation. Prior consent was obtained from all patients before each scanning session, along with the approval of the Cardiff University School of Psychology ethics committee. Acquisition and processing of the data was supported by a Rubicon grant from the NWO (680-50-1527), a Wellcome Trust Investigator Award (096646/Z/11/Z), and a Wellcome Trust Strategic Award (104943/Z/14/Z). We acknowledge the 2017 and 2018 MICCAI Computational Diffusion MRI committees (Francesco Grussu, Enrico Kaden, Lipeng Ning, Jelle Veraart, Elisenda Bonet-Carne, and Farshid Sepehrband) and CUBRIC, Cardiff University (Derek Jones, Umesh Rudrapatna, John Evans, Greg Parker, Slawomir Kusmia, Cyril Charron, and David Linden). 

%
%
\nocite{*}

\bibliography{biblio}
\bibliographystyle{splncs03}

\appendix

\section{Derivation of the Bound in Eq. \ref{eq:IZS}}
\label{sec:app:bound}

This bound is also found in Moyer et al. \cite{NIPS2018_8122}, where it is used in the context of Fair Representations. Again, we reproduce it here for clarity, but the demonstration remain unchanged. All entropic quantities are with respect to $q$ the empirical encoding distribution unless otherwise stated.

From the tri-variate identities of mututal information, we have that $I(z,s) = I(z, x) - I(z, x | s) + I(z, s |x)$. However, the distribution of $z$ is exactly given by $\int q(z|x) dx$ by construction, and thus the distribution of $z$ solely depends on $x$. Thus,
\begin{align}
I(z, s |x) & = H(z|x) - H(z|x,s) = H(z|x) - H(z|x) = 0.
\end{align}
We can then write the following:
\begin{align}
I(z,s) & = I(z, x) - I(z, x | s)\label{eq:fork_in_road}\\
& = I(z, x) - H(x|s) + H(x | z, s) \\
& \leq I(z, x) - H(x|s) - \mathbb{E}_{x,s,z \sim q}[ \log p(x|z,s)] \label{eq:breakdown} \\
& = \mathbb{E}_{z,x}[ \log q(z|x) - \log q(z)] - H(x|s) - \mathbb{E}_{x,s,z\sim q}[ \log p(x|z,s)]\\
& = \mathbb{E}_{x}[~KL[~q(z|x)~\|~q(z)~]~  ] - H(x|s) - \mathbb{E}_{x,s,z\sim q}[ \log p(x|z,s)]. \label{eq:penalty}
\end{align}
This inequality is tight if and only if the variational approximation $p(x|z,s)$ is correct; interpreted in an imaging context, if we cannot perform conditional reconstruction correctly this bound will not be tight.

\end{document}